# Modal analysis of electromagnetic resonators: MAN software expansion to 2D materials and coupled systems

Thomas Christopoulos,[1] Tong Wu,[1,2] and P. Lalanne[1*]

[1] Laboratoire Photonique, Numérique et Nanosciences (LP2N), IOGS-Université de Bordeaux-CNRS, 33400 Talence, France

[2] Institute of Modern Optics, College of Electronic Information and Optical Engineering, Nankai University, Tianjin 300350, China

*philippe.lalanne@institutoptique.fr

**Abstract**

We present Version 9 of the freeware MAN (Modal Analysis of Nanoresonators) [Comput Phys Commun **284**, 108627 (2023)], a software package for computing, normalizing, and exploiting quasinormal modes (QNMs) to analyze the optical response of electromagnetic resonators. This release substantially extends the capabilities of the original software through three major developments. First, it introduces dedicated models and numerical tools for resonators incorporating two-dimensional materials, including graphene, using a surface-conductivity formulation. Second, it implements a coupled-QNM formalism that computes the complex coupling coefficients between resonator modes directly from the QNMs of the individual resonators. Unlike conventional coupled-mode theory, where these coefficients are often introduced phenomenologically or determined by fitting, MAN evaluates them rigorously from first principles. Third, it provides post-processing tools that determine the Fano parameters governing extinction spectra directly from the computed QNM field distributions. Together, these advances broaden the range of nanophotonic systems accessible to MAN while providing new tools for the quantitative interpretation of complex resonant phenomena.

## 1. Introduction

Electromagnetic wave scattering by resonant systems is governed by the excitation of quasinormal modes (QNMs), i.e., the eigenmodes of non-Hermitian systems, by external stimuli [1-3]. QNM theory provides a rigorous and versatile framework to describe excitation, perturbation, and coupling phenomena in such systems [1]. QNMs are characterized by complex resonance frequencies $\tilde{\omega}_m = \omega_m + i\,\gamma_m/2$, reflecting the open and dissipative nature of the system. Here, $\omega_m$ denotes the resonance frequency of mode $m$ and $\gamma_m$ the decay rate, related to the mode lifetime by $\tau_m = 1/\gamma_m$. Under the $\exp\{+i\omega t\}$ time-harmonic convention, $\gamma_m > 0$, leading to temporal decay accompanied by spatial exponential growth. The spatial growth historically raised mathematical and physical concerns, which are now well understood both theoretically [4] and experimentally [5].

In a recent work, we introduced MAN (Modal Analysis of Nanoresonators) [6], a comprehensive software package for computing, normalizing, and exploiting quasinormal modes (QNMs) to analyze the optical response of resonant nanophotonic systems. By enabling QNM-based reduced-order models, MAN significantly lowers computational cost while providing direct physical insight into the mechanisms governing resonant light–matter interactions.

The study of electromagnetic QNMs spans more than three decades. While early works established orthogonality and completeness in simplified geometries [7], they did not resolve the key issue of QNM normalization due to far-field divergence. This challenge was overcome in later studies [8-11], leading to a rigorous and widely accepted normalization formalism [1, 4,12-14]. Combined with the intuitive interpretation offered by modal expansions, these advances have enabled the development of several numerical tools for QNM analysis [6,13,15-17], among which MAN has emerged as a mature and versatile implementation [6].

This article extends Ref. [6] by presenting Version 9 of MAN, which introduces new functionalities and improvements. In FIG. 1 we graphically present the structure of the software, organized in Tutorials, Solvers, COMSOL models, Toolboxes, tests, and the documentation. We use blue fonts to highlight the new functionality. Generally, Version 9 includes three major updates. The respective models, functions, and toolboxes related to the major updates introduced in the present paper are noted in FIG. 1 with underline. First, the framework is extended to efficiently handle dispersive two-dimensional materials, such as graphene, modeled via surface conductivity. Second, a new toolbox is introduced for analyzing electromagnetic coupling in non-Hermitian photonic and plasmonic systems, based on a rigorous formulation that places temporal coupled-mode theory on a first-principles foundation and remains valid beyond weak-coupling regimes. Third, new functions are provided to compute Fano parameters from closed-form expressions, enabling direct and physically grounded analysis of asymmetric spectral lineshapes without fitting procedures.

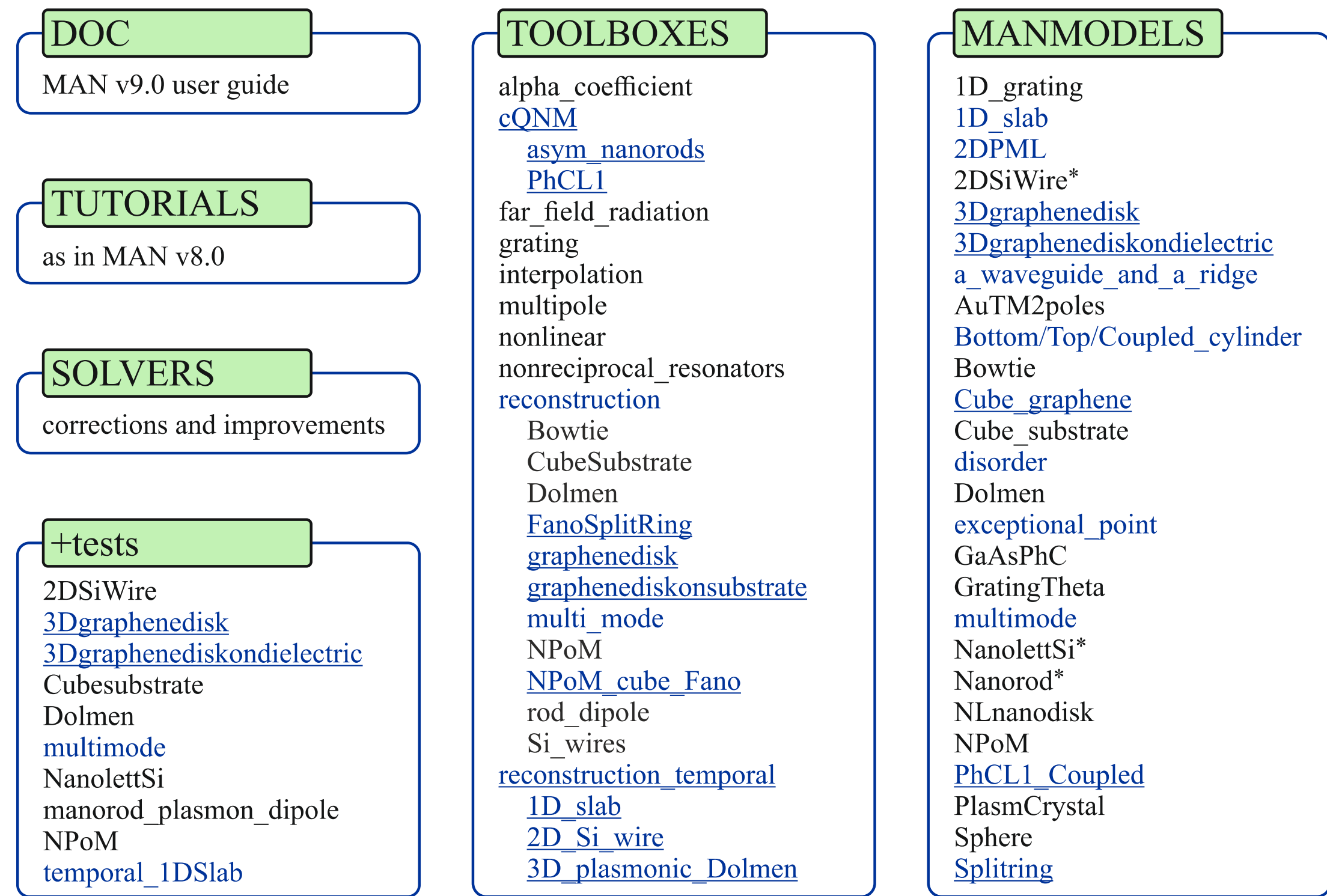

**FIG. 1**. Structure of MAN V9. Black terms already exist in the previous version MAN 8.0. Blue font is used for the new models, functions, and toolboxes introduced in this Version. Underlined models and functions correspond to the three major updates discussed in this work.

These developments are described in the following sections, each accompanied by a brief theoretical overview and fully reproducible numerical examples.

First, motivated by the growing interest in two-dimensional (2D) materials, e.g. graphene, for controlling light–matter interactions [18,19], we extend the QNM solvers and toolboxes to efficiently support dispersive 2D materials. Rather than using bulk equivalents, these materials are modeled via surface current boundary conditions [20,21]. The two solvers implemented in MAN, QNMPole [11] and QNMEig [14], are accordingly extended to compute QNMs of systems described by surface conductivity. QNMPole naturally incorporates 2D materials within its general framework for arbitrary geometries and dispersions, while QNMEig relies on the eigenmode solver of COMSOL and treats dispersive media by linearizing the eigenvalue problem through the introduction of auxiliary fields [14]. The reconstruction toolbox is also updated to properly handle structures incorporating 2D materials.

Second, we introduce a new toolbox for analyzing electromagnetic coupling between non-Hermitian photonic and plasmonic resonators. Based on a recent theoretical framework [22], the toolbox provides closed-form expressions for the dissipative coupling coefficients and enables direct computation of the supermodes of coupled systems from the QNMs of the individual resonators. In contrast to conventional temporal coupled-mode theory (TCMT) [23-25], where the coupling coefficients are often introduced phenomenologically or determined by fitting, the present approach evaluates all dynamical parameters—including coupling coefficients, eigenfrequencies, and source terms—analytically from first principles. It also extends the temporal toolbox of MAN Version 8 [6,26,27] beyond the weak-coupling and high-Q regimes in which TCMT is generally applicable.

Third, we introduce new functions for computing the Fano parameters governing asymmetric extinction spectra [28]. Rather than extracting these parameters by fitting spectral line shapes, the toolbox evaluates them analytically from closed-form expressions involving the QNMs and their overlap with the incident field. Once the complex eigenfrequencies have been computed, the contribution of each QNM to the extinction spectrum is determined uniquely, providing a physically transparent and predictive interpretation of Fano resonances [29].

## 2. Two-dimensional materials

QNMs are the eigensolutions of the source-free curl Maxwell's equations

$$\nabla \times \tilde{\mathbf{E}}_m(\mathbf{r}) = -i\tilde{\omega}_m\bar{\bar{\mu}}(\mathbf{r},\tilde{\omega}_m)\tilde{\mathbf{H}}_m(\mathbf{r}), \tag{1a}$$

$$\nabla \times \tilde{\mathbf{H}}_m(\mathbf{r}) = i\tilde{\omega}_m\bar{\bar{\varepsilon}}(\mathbf{r},\tilde{\omega}_m)\tilde{\mathbf{E}}_m(\mathbf{r}) + \bar{\bar{\sigma}}_s(\mathbf{r},\tilde{\omega}_m)\tilde{\mathbf{E}}_m(\mathbf{r}), \tag{1b}$$

+ outgoing boundary conditions.

The previous equations are written in the most general form supported by MAN, covering dispersive, anisotropic, and non-reciprocal materials. Earlier versions of MAN were limited to bulk systems with vanishing surface conductivity ($\bar{\bar{\sigma}}_s = 0$). A key update in the new version is the ability to model non-zero, dispersive surface conductivities [30]. Both QNMPole and QNMEig have been upgraded to correctly compute and normalize QNMs for $\bar{\bar{\sigma}}_s(\mathbf{r}) \neq 0$, and new functions have been introduced in the reconstruction toolbox to compute the new QNM excitation coefficients arising from $\bar{\bar{\sigma}}_s(\mathbf{r})$.

### 2.1. QNM theory for two-dimensional materials

Generally, a 2D material can be handled by an equivalent anisotropic bulk medium [20,21,31], a case that the previous version of MAN is able to handle. However, to get an accurate representation, the thickness of the 2D material should be typically below 1 nm, inevitably requiring quite small discretization and therefore significantly increasing the computational burden.

Alternatively, 2D materials can be treated as surface current boundary conditions [20,21,30] by imposing a discontinuity in the tangential magnetic field. More physically intuitive, this treatment is also much more computationally efficient, requiring the discretization of a surface rather than an ultra-thin volume.

For modeling a 2D material as a surface boundary, we introduce the *surface* conductivity tensor $\bar{\bar{\sigma}}_s(\mathbf{r},\omega)$ as an infinitesimally thin version of the bulk conductivity $\bar{\bar{\sigma}}(\mathbf{r},\omega) = \bar{\bar{\sigma}}_s(\mathbf{r},\omega)\delta(\mathbf{r}-\mathbf{r}_s)$, with $\mathbf{r}_s$ denoting the position of the 2D material. Computationally, the 2D material can be treated as a surface current density $\mathbf{J}_s$ in the tangential magnetic field boundary condition [20,21]

$$\hat{\mathbf{n}} \times (\mathbf{H}_2 - \mathbf{H}_1) = \mathbf{J}_s \equiv \bar{\bar{\sigma}}_s(\mathbf{r},\omega)\mathbf{E}_1, \tag{2}$$

with 2 and 1 labelling the two media surrounding the 2D material (see Fig. 2) and $\hat{\mathbf{n}}$ denoting the normal vector to the interface, pointing towards medium 2. As 2D materials are generally isotropic in-plane (in the absence of external static magnetic bias) and do not interact with the out-of-plane electric field component, it is convenient to define the in-plane parallel electric field as $\mathbf{E}_{\parallel}(\mathbf{r},\omega) = \hat{\mathbf{n}} \times \mathbf{E}(\mathbf{r},\omega) \times \hat{\mathbf{n}}$. Then, we can simplify the product $\bar{\bar{\sigma}}_s(\mathbf{r},\omega)\mathbf{E}(\mathbf{r},\omega) = \sigma_s(\mathbf{r},\omega)\mathbf{E}_{\parallel}(\mathbf{r},\omega)$ using the scalar $\sigma_s(\mathbf{r},\omega)$, defined as the only independent element of $\bar{\bar{\sigma}}_s(\mathbf{r},\omega)$.

The computational burden of treating a 2D material as either a surface boundary or an equivalent bulk medium can be better understood by comparing the two representations, see Fig. 2. There, a single graphene ribbon is either modeled as a surface current density $\mathbf{J}_s(\mathbf{r})$ or an equivalent very thin layer with an anisotropic permittivity $\bar{\bar{\varepsilon}}_{3\mathrm{D}} = \varepsilon_0 \mathrm{diag}(\varepsilon_{\mathrm{eq}}, \varepsilon_{\mathrm{eq}}, 1)$, with $\varepsilon_{\mathrm{eq}} = 1 - i\sigma_s/(\omega\varepsilon_0 d)$ and $d$ being the thickness of the layer, typically around 1 nm.

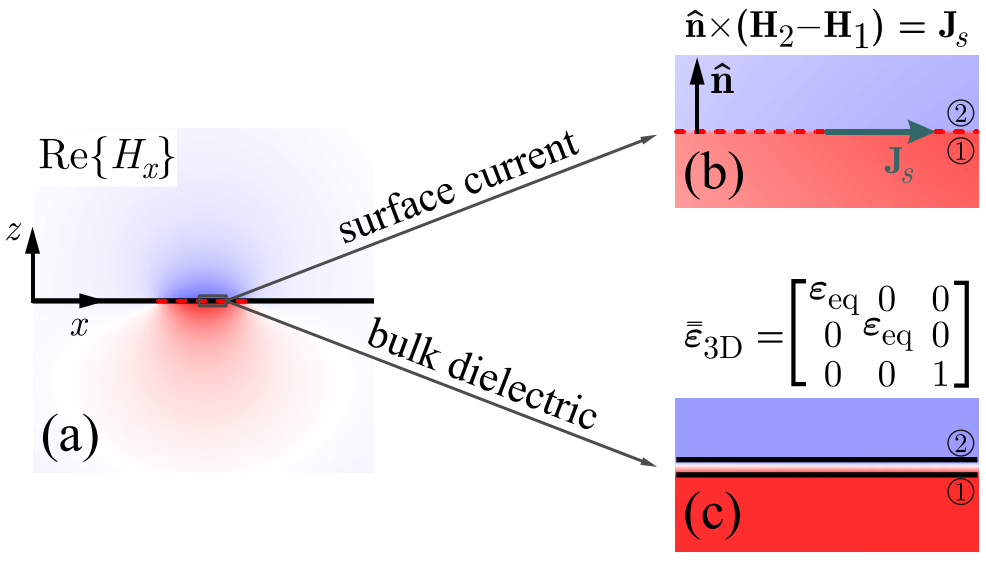

**FIG. 2**. (a) Field plot of the tangential magnetic field in the $xz$-plane of a single graphene ribbon. (b,c) Zoomed-in field plot, highlighting the differences of modeling graphene either with a surface current boundary condition (b) or with an equivalent bulk anisotropic thin layer (c).

Macroscopically, the surface and the effective bulk representations are equivalent, leading to the same field distribution of a graphene plasmonic mode [Fig. 2(a)]. As dictated by Eq. (2), the surface current density $\mathbf{J}_s$ induces a discontinuity in the tangential magnetic field [Fig. 2(b)]. In the equivalent bulk dielectric representation [Fig. 2(c)], the tangential magnetic field in the two interfaces should be continuous and therefore the change of sign is induced inside the artificial volume. To capture this large field change, a dense discretization inside the equivalent bulk layer is eminent.

The surface conductivity of the 2D material may be dispersive, e.g., Drude [30], Drude-Lorentz [32], or complex-conjugate Debye [33]. Essentially, the treatment is the same and in the current version of MAN, we focus on the simplest case, i.e., Drude,

$$\sigma_s(\mathbf{r},\omega) = -i\frac{\sigma_0(\mathbf{r})}{\omega - i\gamma}, \tag{3}$$

with $\sigma_0(\mathbf{r})$ a frequency independent term that characterizes the 2D material and $\gamma$ its decay rate.

The Drude model of Eq. (3) is usually sufficient to describe 2D materials with surface conductivities dominated by interband interactions in the mid-infrared, like graphene [34], MXenes [35], or some doped Xenes (e.g., silicene, germanene) [36], i.e., single or few-layer exfoliants of elements of the 14th (IVA) group. They all have metallic-like behavior, supporting surface plasmons. On the other hand, 2D materials with surface conductivities dominated by intraband interactions which have, therefore, more semiconductor-like responses due to the formed excitons, are usually described by a (multipole) Drude-Lorentz model. Such materials are single or few-layer transition metal dichalcogenides (TMDCs) [37], black phosphorus [38], undoped Xenes (e.g., antimonene, bismuthene) [36], or 2D topological insulators (e.g., $Bi_2Se_3$) [39]. Although we do not include a tutorial model for such materials, their description with the framework has already been presented in the literature [32] and can be easily extended from the presented models. As briefly discussed in the introduction, depending on the solver, the computational description of 2D materials in MAN is performed differently. In QNMPole, the 2D material is introduced by a *surface current density* boundary condition, imposing the connection between the surface current $\mathbf{J}_s(\mathbf{r},\omega)$ and the tangential electric field $\mathbf{E}_\parallel(\mathbf{r},\omega)$ at frequency $\omega$, i.e., $\mathbf{J}_s(\mathbf{r},\omega) = \sigma_s(\mathbf{r},\omega)\mathbf{E}_\parallel(\mathbf{r},\omega)$.

For the implementation in QNMEig, an auxiliary field should be used. For Drude dispersion, this field is $\tilde{\mathbf{J}}_m(\mathbf{r}) = -i\sigma_s(\mathbf{r},\tilde{\omega}_m)\tilde{\mathbf{E}}_{\parallel,m}(\mathbf{r})$, which is deliberately chosen to be different from the surface current term $\mathbf{J}_s(\mathbf{r},\omega)$ by a $-i$ prefactor. $\tilde{\mathbf{J}}_m(\mathbf{r})$ should be included in the finite-element formulation that MAN  resorts to. Specifically, by applying the Galerkin technique [40], we can find the appropriate weak-form expression of the Helmholtz equation for the $m$-th QNM,

$$\iiint_\Omega \left[(\nabla\times\mathbf{E}')\cdot\left(\bar{\bar{\mu}}^{-1}\nabla\times\tilde{\mathbf{E}}_m\right) - \tilde{\omega}_m^2\mathbf{E}'\cdot\left(\bar{\bar{\varepsilon}}\tilde{\mathbf{E}}_m\right)\right]\mathrm{d}V - \iint_S \tilde{\omega}_m\mathbf{E}_\parallel'\cdot\tilde{\mathbf{J}}_m\mathrm{d}S + \mathrm{B.C.} = 0. \tag{4}$$

The notation B. C. implies appropriate termination boundary conditions and vector quantities with a prime denote test functions. $\Omega$ represents the full simulation domain, including the PML region and $S$ is the area where the surface conductivity is non-zero. The integration for this term might as well extend inside PML. This weak form should be accompanied by an additional weak expression for the auxiliary field, so that the emerging eigenvalue problem is linearized [14,30]. In the case of Drude dispersion considered here, we find

$$\iint_S \left[\tilde{\omega}_m\mathbf{J}'\cdot\tilde{\mathbf{J}}_m + \mathbf{J}'\cdot(-i\gamma\tilde{\mathbf{J}}_m) + \mathbf{J}'\cdot\left(\sigma_0\tilde{\mathbf{E}}_{\parallel,m}\right)\right]\mathrm{d}S = 0. \tag{5}$$

The extra surface term in Eq. (4) involving $\tilde{\mathbf{J}}_m$ is introduced in COMSOL weak formalism via a *surface weak contribution* term. Equation (5) is included as an additional *weak form boundary PDE*, coupled with the main physics through the weak form of the Helmholtz equation. Then, Eqs. (4) and (5), which form a quadratic eigenproblem, are solved concurrently with the eigensolver of COMSOL to compute the QNMs.

In terms of QNM normalization, the inclusion of a dispersive 2D material has an impact and should be treated consistently. Therefore, QNMs are normalized such that [30]

$$\iiint_\Omega \left[\tilde{\mathbf{E}}_m(\mathbf{r})\cdot\frac{\partial\{\omega\bar{\bar{\varepsilon}}(\mathbf{r},\omega)\}}{\partial\omega}\tilde{\mathbf{E}}_m(\mathbf{r}) - \tilde{\mathbf{H}}_m(\mathbf{r})\cdot\frac{\partial\{\omega\bar{\bar{\mu}}(\mathbf{r},\omega)\}}{\partial\omega}\tilde{\mathbf{E}}_m(\mathbf{r})\right]\mathrm{d}V$$

$$-\iint_S \tilde{\mathbf{E}}_{\parallel,m}(\mathbf{r})\cdot i\frac{\partial\sigma_s(\mathbf{r},\omega)}{\partial\omega}\tilde{\mathbf{E}}_{\parallel,m}(\mathbf{r})\mathrm{d}S = 1. \tag{6}$$

The first two terms of Eq. (6) are well established in the QNM theory literature, while the last term captures the contribution of 2D materials. All derivatives in Eq. (6) are calculated at the complex QNM frequency $\tilde{\omega}_m$. Finally, we shall note that, as Eq. (6) directly dictates, throughout the article the pair $[\tilde{\mathbf{E}}_m(\mathbf{r}), \tilde{\mathbf{H}}_m(\mathbf{r})]$ denotes the electric and magnetic fields of *normalized* QNMs.

The contribution of the 2D materials in the expansion coefficient is independent of the contribution of bulk materials and one may write $\alpha_m(\omega) = \alpha_{m,\mathrm{3D}}(\omega) + \alpha_{m,\mathrm{2D}}(\omega)$ [32]. The calculation of $\alpha_{m,\mathrm{3D}}(\omega)$ is well documented in Ref. [1] and nothing new is added in MAN  V9 on that; we include the expression to calculate $\alpha_{m,\mathrm{3D}}(\omega)$ below for convenience, along with the newly introduced expression for $\alpha_{m,\mathrm{2D}}(\omega)$ [30]

$$\alpha_{m,\mathrm{3D}}(\omega) = \iiint_{\Omega_{\mathrm{res}}} \left[\varepsilon_b(\mathbf{r},\omega) - \varepsilon_\infty(\mathbf{r}) + \Delta\varepsilon(\mathbf{r},\tilde{\omega}_m)\frac{\tilde{\omega}_m}{\tilde{\omega}_m-\omega}\right]\tilde{\mathbf{E}}_m(\mathbf{r})\cdot\mathbf{E}_b(\mathbf{r},\omega)\mathrm{d}V, \tag{7a}$$

$$\alpha_{m,\mathrm{2D}}(\omega) = \iint_{S_{\mathrm{res}}} -i\sigma_s(\mathbf{r},\tilde{\omega}_m)\frac{1}{\tilde{\omega}_m-\omega}\tilde{\mathbf{E}}_{\parallel,m}(\mathbf{r})\cdot\mathbf{E}_b(\mathbf{r},\omega)\mathrm{d}S. \tag{7b}$$

$\mathbf{E}_b(\mathbf{r},\omega)$ denotes the background field, i.e., the incident field plus any reflection by a substrate, $\varepsilon_b(\mathbf{r},\omega)$ is the background permittivity, $\Delta\varepsilon(\mathbf{r},\omega) = \varepsilon(\mathbf{r},\omega) - \varepsilon_b(\mathbf{r},\omega)$ is the permittivity difference defining the resonator volume $\Omega_{\text{res}}$, $S_{\text{res}}$ denotes the area of the resonator with non-vanishing $\sigma_s$, and $\varepsilon_\infty(\mathbf{r}) = \varepsilon(\mathbf{r},\omega\to\infty)$. In the new version of MAN , a newly introduced script performs the calculation of $\alpha_{m,\text{2D}}(\omega)$.

In the same spirit, the contribution of bulk and sheet materials in the extinction and absorption cross-sections can be separated, i.e., $\sigma(\omega) = \sigma_{\text{3D}}(\omega) + \sigma_{\text{2D}}(\omega)$. Again, the bulk case is well documented in Ref. [1], while for the 2D materials we use the expressions

$$\sigma_{\text{ext,2D}}(\omega) = -\frac{1}{2I_0}\iint_{S_{\text{res}}} \text{Im}\left\{-i\sigma_s(\mathbf{r},\omega)\mathbf{E}_{\parallel,\text{tot}}(\mathbf{r},\omega)\cdot\mathbf{E}^*_{\parallel,b}(\mathbf{r},\omega)\right\}\text{d}S, \tag{8a}$$

$$\sigma_{\text{abs,2D}}(\omega) = \frac{1}{2I_0}\iint_{S_{\text{res}}} \text{Re}\{\sigma_s(\mathbf{r},\omega)\}\left|\mathbf{E}_{\parallel,\text{tot}}(\mathbf{r},\omega)\right|^2\text{d}S. \tag{8b}$$

The usefulness of this decomposition will become evident in Sec. 2.2, where we demonstrate results using the upgraded reconstruction toolbox for systems with 2D materials included in MAN V9. As with the bulk case, several expansions of $\sigma_s(\mathbf{r},\omega)\tilde{\mathbf{E}}_{\parallel,\text{tot}}(\mathbf{r},\omega)$ can be considered for the reconstruction [6,41], but in MAN V9 we explicitly adopt the formulation $\sigma_s(\mathbf{r},\omega)\tilde{\mathbf{E}}_{\parallel,\text{tot}}(\mathbf{r},\omega) = \sum_m \alpha_m(\omega)\sigma_s(\mathbf{r},\tilde{\omega}_m)\tilde{\mathbf{E}}_{\parallel,m}(\mathbf{r})$, which seems to be the one that gives the fastest and most accurate convergence, considering a minimal number of modes.

### 2.2. COMSOL models and MATLAB toolboxes demonstration

MAN V9 update includes three models that show how one can compute and normalize the QNMs of systems containing 2D materials, modeled exclusively with surface conductivities. The QNMs for all three models are computed using the newly developed COMSOL models, employing either the QNMPole or QNMEig solvers. The extinction and absorption cross-sections are reconstructed using the newly built MATLAB functions in the reconstruction toolbox.

We first present the model **graphenedisk.mph**, consisting of a single graphene disk of 40 μm radius on a dielectric substrate [see Fig. 3(a) for a schematic]. The disk supports graphene surface plasmons (GSPs) that are tightly confined in the graphene-dielectric interface, such as the dipole-like QNM of which the normalized $\text{Re}\{\tilde{E}_z\}$ field distribution is depicted in Fig. 3(a). We focus on the reconstruction of its response using Eqs. (7) and (8). These equations have been implemented in newly introduced MATLAB functions in the reconstruction toolbox. In Fig. 3(b,c), we show the extinction and absorption cross-sections for a normally incident plane wave reconstructed using the new functions, considering either only the dominant dipolar QNM or a set of 100 modes including both QNMs and PML modes. The results are compared with reference scattering simulation at the real frequency (red dots), again performed with COMSOL using identical mesh and modeling graphene as a surface current boundary condition with dispersive conductivity.

As the QNMs supported by the graphene disk are spectrally well separated, the single QNM reconstruction in the vicinity of the dipolar mode resonance frequency ~0.6 THz is already accurate. Perfect agreement is found for the 100 QNMs reconstruction result. Note that this single QNM is computed in about two minutes with QNMEig by adopting the surface auxiliary field formulation and approximately 400 000 degrees of freedom (DoFs). For the equivalent representation of graphene as a 1-nm-thick bulk material, the equivalent graphene layer must be finely discretized along the $z$ direction to capture the continuous yet strongly varying field inside [see Fig. 2(c)]. As a result, the number of DoFs increases to approximately 2.8 million, leading to a substantial rise in memory requirements and computational cost. Consequently, the computation time per QNM increases by roughly a factor of 20, reaching up to 40 minutes.

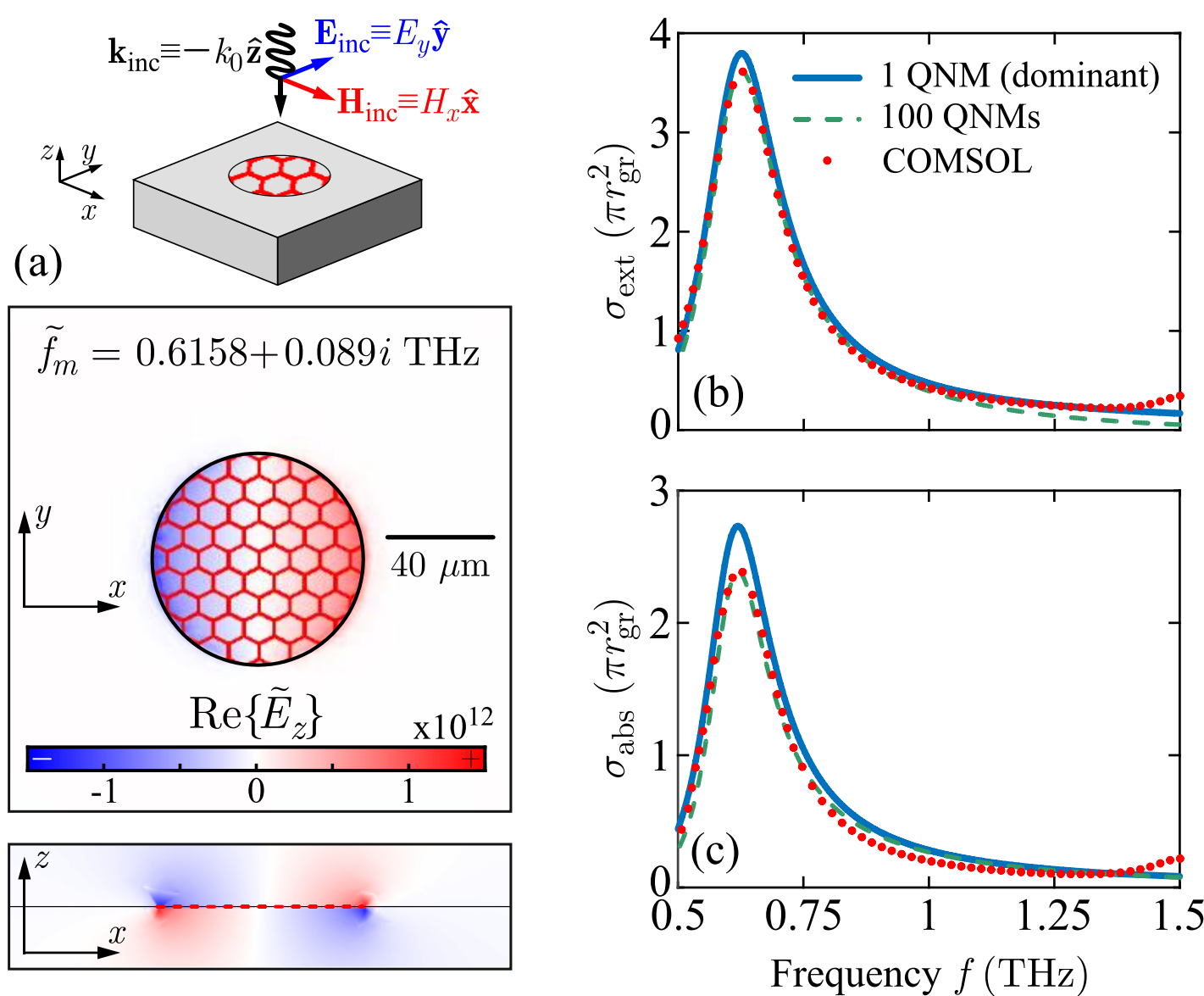

**FIG. 3**. Reconstruction toolbox results for a graphene disk on a glass substrate. Material properties for graphene: $\sigma_0 = 2.35 \times 10^{10}$ S/m and $\gamma = 10^{12}$ rad/s, and for the substrate: $\varepsilon_r = 2.25$. (a) Schematic diagram of the graphene disk on a substrate and normalized field plots (Re$\{\tilde{E}_z\}$) of its dipole-like QNM in the $xy$ and $xz$ planes. (b) Extinction cross-section and (c) absorption cross-section near the resonance frequency of the dipolar QNM. Accurate reconstruction is achieved even when only the dipolar QNM is considered.

We also examine a variation of the previous system, **graphenediskondielectric.mph**. Now, a 45-μm-radius graphene disk lies on top of an 110-nm-height dielectric rod. For the rod, we choose a material with high permittivity in the THz band to showcase the interactions between the plasmonic mode of the graphene disk and the photonic mode of the rod. Therefore, we use $LiTaO_3$ with a relative permittivity around 40 in the far infrared and very weak (negligible) dispersion. The simulated cavity system is shown in Fig. 4(a). The dotted curve in Fig. 4(b) displays its extinction cross section $\sigma_{\rm ext}$, obtained from real-frequency spectral scattering simulations (dots). Solid curves with different colors correspond to contributions of different QNMs to $\sigma_{\rm ext}$. The newly built functions in the reconstruction toolbox combine the expansion coefficient and scattering cross-section contributions from both bulk (3D) and 2D materials. This capability is crucial in the present study, as the resonators involve both 2D and 3D structures. Besides, these functions allow for disparate the contributions of different structures, which is important for understanding how the modes of the components are hybridized [42].

To demonstrate this, we consider QNM2 [brick red curve in Fig. 4(b)] and show in Fig. 4(c) its contribution $\sigma_2$ to the extinction cross-section $\sigma_{\rm ext}$. This contribution can be decomposed as $\sigma_2(\omega) = \sigma_{2,\rm 3D}(\omega) + \sigma_{2,\rm 2D}(\omega)$. We observe that $\sigma_{2,\rm 2D}(\omega)$ is negative, indicating that the 2D material reduces the overall extinction. This negative contribution implies that the 2D material suppresses near-field light-matter interaction through destructive interference. In other words, the incident plane wave induces opposing effective polarizations in the 2D and 3D media, leading to partial cancellation between their responses. This cancellation is caused by the hybridization between the QNM of the graphene disk and that of the dielectric rod. The hybridization is illustrated by the inset in Fig. 4(b). The two plots on the right-hand side depict the Re$\{\tilde{E}_z\}$ field component of the plasmonic (top) and photonic (bottom) mode in the absence of the rod or graphene, respectively. The plot on the left-hand side shows the hybridized QNM2. Evidently, the two original modes interfere destructively and the anti-symmetric supermode exhibits a lower excitation cross-section. One can engineer this interaction to control, e.g., the excitation efficiency of such a hybrid mode [43].

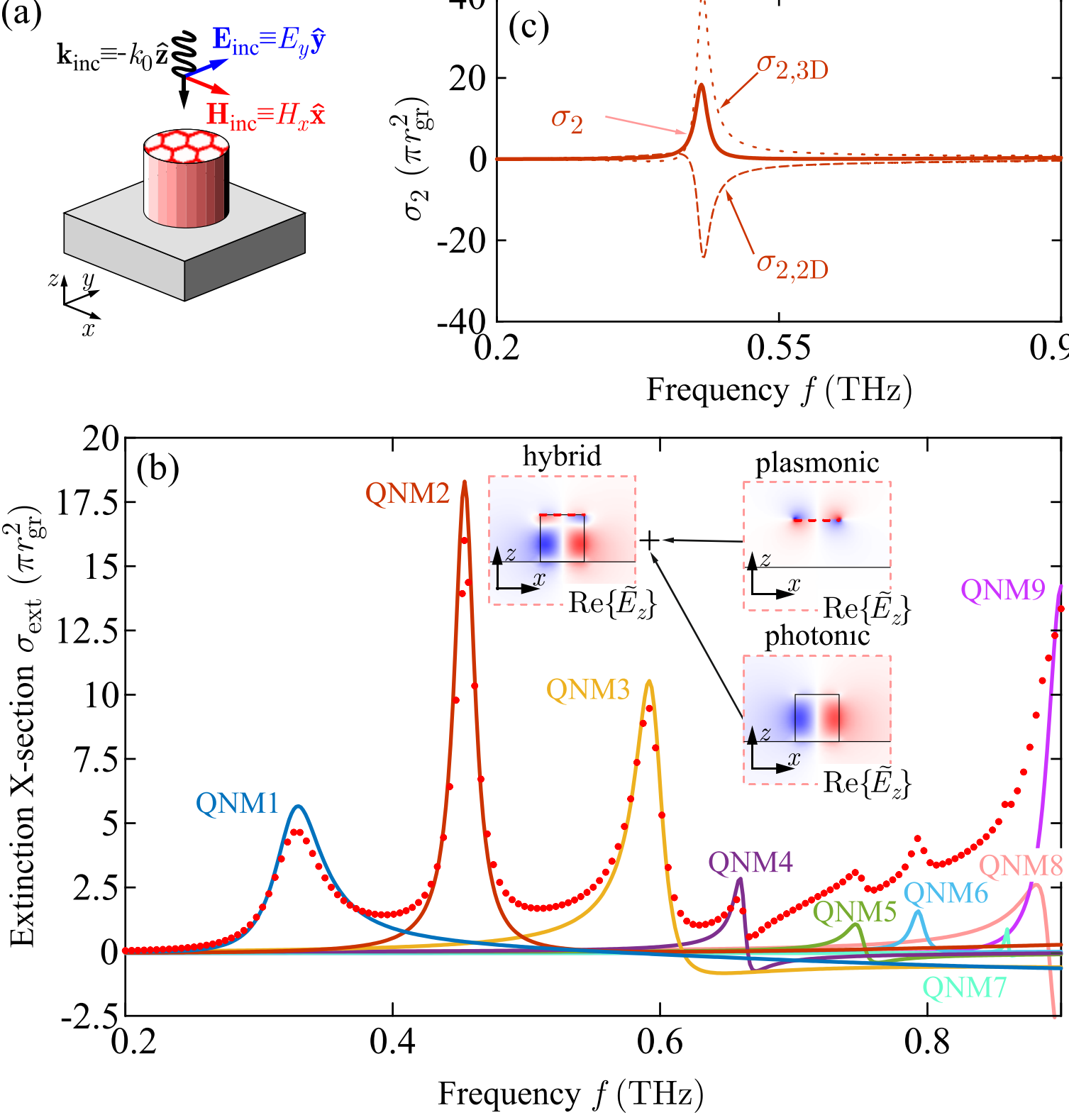

**FIG. 4**. Reconstruction toolbox results for a graphene disk on top of a high-index dielectric rod. Material properties for graphene: $\sigma_0 = 11.8 \times 10^{10}$ S/m and $\gamma = 10^{12}$ rad/s, for the $LiTaO_3$ rod: $\varepsilon_r = 41.35$, and for the substrate: $\varepsilon_r = 2.25$. (a) Schematic diagram of the considered structure: A 45-µm-radius graphene disk lies on top of a 110-nm-height dielectric rod residing on a substrate. The system is excited by a normally incident plane wave. (b) Extinction cross-section and contributions of various QNMs. Inset: QNM2 is caused by the hybridization between the QNM of the rod and that of the graphene disk. Plots on the right side show the plasmonic modes supported by graphene in the absence of the rod (top) and the photonic mode supported by the rod in the absence of graphene (bottom). The plot on the left shows the hybrid mode. (c) Decomposition of the QNM2 contribution to extinction cross-section in its bulk and surface parts. The bulk (photonic) part of the mode generally dominates, while the surface (plasmonic) part acts in opposition, thereby suppressing the observable response of the system.

Finally, the more complex model **Cube_graphene.mph** is discussed, where we demonstrate the extreme light confinement in the nanogap between a metallic nanocube and graphene. Technically, **Cube_graphene.mph** is an example to show how one can comprehensively build a COMSOL model consisting of both dispersive bulk *and* 2D materials. We include COMSOL models built in both QNMPole and QNMEig solvers in MAN V9, with the latter being of special importance as it combines auxiliary fields with bulk and surface terms. This system is also unique in terms of physics as it achieves extreme mode confinement not for visible light but rather for electromagnetic waves in the far-infrared [44]. The mode volume of the cavity is in the order of $10^{-7}$ cubic wavelengths, quite a few orders of magnitude smaller than the mode volume of the same nanocube over a metallic substrate with an equally thin spacer in the visible [44]. The analyzed structure is depicted in the top-left inset of Fig. 5 and is a gold cube of 75 nm side length on a silica substrate. Between the cube and the substrate lies a 3 nm thick dielectric spacer and below an infinitesimally thin graphene monolayer.

In Fig. 5, the complex frequency plane computed with the QNMEig solver after a single run is displayed. All depicted QNMs of the nanogap possess extremely low mode volumes, as a result of the supported graphene surface plasmons being further confined by the gold nanocube. To demonstrate their plasmonic nature, we include their dominant (normal) field component $\mathrm{Re}\{\tilde{E}_z\}$, in both $xy$- (middle of the nanogap) and $xz$-planes, see the bottom three insets.

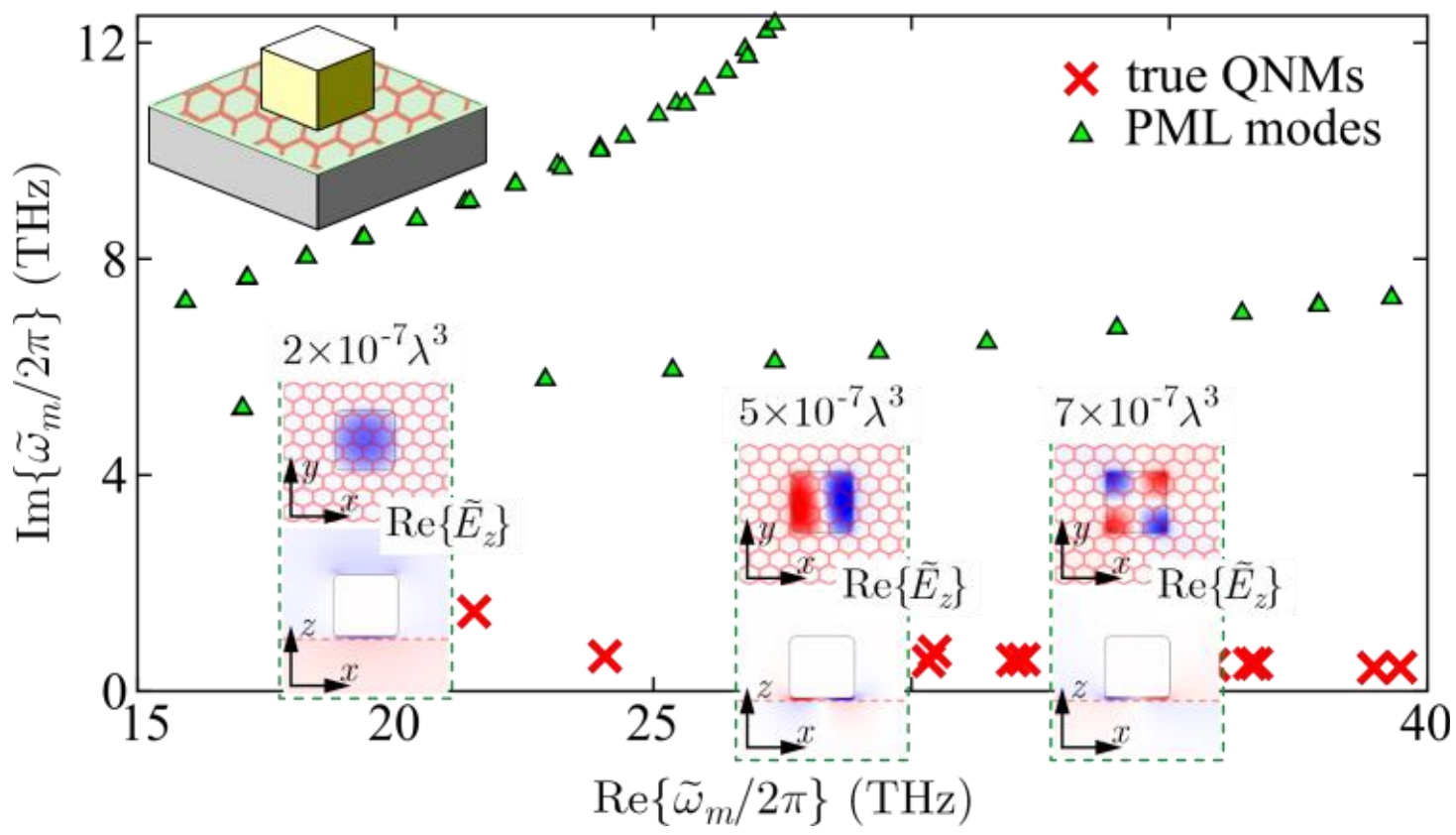

**FIG. 5**. QNMEig results for a gold nanocube on top of a graphene monolayer with a 3-nm-thick dielectric layer between them. Material properties for gold: $\omega_p = 1.37 \times 10^{16}$ rad/s and $\gamma = 7.64 \times 10^{13}$ rad/s, for graphene: $\sigma_0 = 7.06 \times 10^{10}$ S/m and $\gamma = 2 \times 10^{12}$ rad/s, for the nanogap: $\varepsilon_r = 2 - i0.05$, and for the substrate: $\varepsilon_r = 2.25$. The complex plane is shown, with true QNMs (red X markers) and PML modes (green triangle markers). Insets: $\mathrm{Re}\{\tilde{E}_z\}$ (normal) field component of a few QNMs in the middle of the nanogap (top) and in appropriate $xz$-cuts (bottom). All the supported modes are plasmonic, with ultra-low mode volumes and emerge due to the interaction between graphene surface plasmons and the gold nanocube.

## 3. Coupled-QNM toolbox

A central strategy in optical resonance design is the deliberate coupling and hybridization of individual resonances to tailor spectral responses. In nanophotonics, devices involving modal coupling are most analyzed using temporal coupled-mode theory (TCMT). TCMT is a phenomenological framework in which each resonant mode is represented as a harmonic oscillator [23,24], enabling a clear description of how the eigenfrequencies and optical responses of a coupled system emerge from interactions among individual modes. Since its introduction, TCMT has been extensively developed and widely employed as a powerful tool for the analysis and engineering of both linear and nonlinear photonic systems [25,27]. However, TCMT is only valid for cavities with high quality factors, i.e., for Hermitian or quasi-Hermitian systems, and limited in the weak coupling regime [23-25,45]. Moreover, mode couplings and interactions with external excitation fields are generally characterized by phenomenological coefficients extracted from experimental or numerical fitting, limiting both predictive capability and physical transparency.

Recently, a rigorous coupled quasinormal mode (cQNM) framework based on coupled-QNM theory has been developed to analyze dissipative coupling in photonic and plasmonic resonators [22]. This framework overcomes the limitations of conventional TCMT and remains valid for both strongly coupled and highly dissipative resonators. Importantly, excitation and coupling coefficients are obtained from closed-form expressions and do not rely on fitting procedures.

The implementation of the cQNM framework is straightforward. Once the QNMs of the individual resonators are computed, the relevant coupling coefficients are obtained from closed-form expressions. The eigenfrequencies of the resulting supermodes are then determined by diagonalizing a low-dimensional matrix, enabling efficient and accurate prediction of the optical response of the coupled-resonators system.

In MAN V9, we include a **cQNM** toolbox dedicated to the coupling between two resonators labeled as "$A$" and "$B$". Extension to multiple resonators is straightforward. The central result of the cQNM framework is the evolution equation that governs the dynamics of the coupled resonators:

$$-i\frac{\mathrm{d}}{\mathrm{d}t}\begin{bmatrix}\boldsymbol{\beta}^A(t)\\ \boldsymbol{\beta}^B(t)\end{bmatrix} = \begin{bmatrix}\tilde{\boldsymbol{\Omega}}^A & \tilde{\boldsymbol{\kappa}}^{AB}\\ \tilde{\boldsymbol{\kappa}}^{BA} & \tilde{\boldsymbol{\Omega}}^B\end{bmatrix}\begin{bmatrix}\boldsymbol{\beta}^A(t)\\ \boldsymbol{\beta}^B(t)\end{bmatrix} + i\begin{bmatrix}\mathbf{0} & \tilde{\mathbf{g}}^{AB}\\ \tilde{\mathbf{g}}^{BA} & \mathbf{0}\end{bmatrix}\frac{\mathrm{d}}{\mathrm{d}t}\begin{bmatrix}\boldsymbol{\beta}^A(t)\\ \boldsymbol{\beta}^B(t)\end{bmatrix} + \begin{bmatrix}\mathbf{F}^A(t)\\ \mathbf{F}^B(t)\end{bmatrix}, \quad (9)$$

where $\boldsymbol{\beta}^A$ and $\boldsymbol{\beta}^B$ are column vectors representing the modal excitation coefficients, $\tilde{\boldsymbol{\Omega}}^A$ and $\tilde{\boldsymbol{\Omega}}^B$ are diagonal matrices of the complex resonance frequencies of the individual resonators, and $\mathbf{F}^A$ and $\mathbf{F}^B$ are the driving source terms. Finally, $\tilde{\boldsymbol{\kappa}}^{AB}$, $\tilde{\boldsymbol{\kappa}}^{BA}$, $\tilde{\mathbf{g}}^{AB}$, $\tilde{\mathbf{g}}^{BA}$ are coupling matrices determining the coherent and dissipative couplings between two resonators. Although the new formulation is reminiscent of the conventional evolution equation of TCMT [25,45], in cQNM theory, coupling is not only performed via the coupling coefficients of the amplitudes $\boldsymbol{\beta}(t)$ but also involves their time derivatives $\mathrm{d}\boldsymbol{\beta}/\mathrm{d}t$.

The closed-form expressions for all coupling matrices and source terms are provided in [22] and in the user guide of the **cQNM** toolbox. Their computation just requires performing overlap integrals between QNMs and/or between QNMs and the incident field. Notably, once the QNMs of the individual resonators have been computed, all quantities appearing in Eq. (9) can be readily obtained irrespective of the relative positions of the resonators, enabling rapid and efficient exploration of the design parameter space.

The **cQNM** toolbox introduced in MAN V9 provides several examples and functions that showcase how to compute the coupling matrices in Eq. (9) and find the eigenfrequencies of the coupled system by solving the eigenequation, i.e., Eq. (9) without the source terms, $\mathbf{F}^A$ and $\mathbf{F}^B$. As an illustrative example, Fig. 6 examines the interaction between a metallic stripe cavity supporting Fabry-Perot-like modes and an ultrasmall plasmonic (Drude) resonator. The interaction is mediated by a metallic stripe waveguide, as shown schematically in Fig. 6(a). Two independent COMSOL models are constructed in MAN V9 to compute the QNMs of the two individual resonators. In this example, only a single QNM is retained for each cavity, reducing the coupled problem of Eq. (9) (without the source terms) to a $2 \times 2$ eigenvalue equation. Depending on the position and shape of the small Drude resonator, we observe either attraction [Fig. 6(b), Position $P_1$] or repulsion [Fig. 6(c), Position $P_2$] of the two supermodes. The results obtained with the **cQNM** toolbox utilizing the QNMs of individual (uncoupled) cavities (solid lines) practically coincide with COMSOL simulations of the full, coupled system (markers).

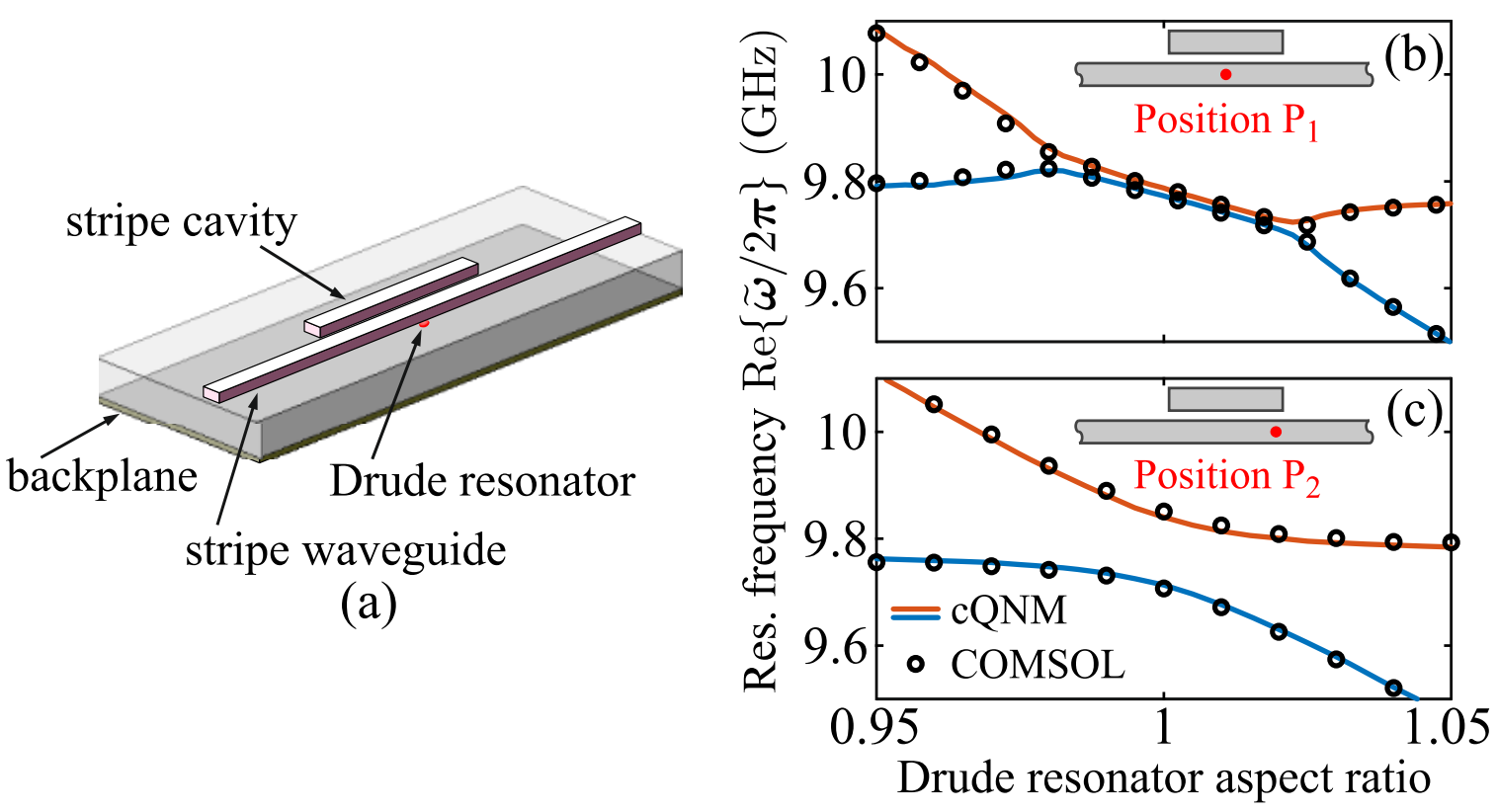

**FIG. 6**. Numerical results obtained with the **cQNM** toolbox for a stripe cavity (**cQNM_waveguide.mph** model) interacting with an ultra-small plasmonic (Drude) nanoresonator (**cQNM_sphere.mph** model) via a stripe waveguide. Materials properties for the stripes: $\omega_p = 745 \times 10^9$ rad/s , $\gamma = 0.6 \times 10^9$ rad/s , and $\varepsilon_\infty = 1$ , for the Drude nanoresonator: $\omega_p = 745 \times 10^9$ rad/s, $\gamma = 3.77 \times 10^9$ rad/s, and $\varepsilon_\infty = 3.38$, and for the gap permittivity: $\varepsilon_r = 3.38$ . (a) Schematic of the examined structure. Stripe cross-section is $2.5 \times 2.5$ mm$^2$ and its length is 40 mm. Drude cavity is an ellipsoid of 0.61 mm length along its one and $0.61 \times$ Aspect ratio along its other axis. Dielectric spacer is 1.5 mm thick. (b) Level attraction and (c) level repulsion of the two supermodes with respect to the Drude nanocavity position and shape. Solid curves correspond to the results obtained by **cQNM** toolbox using the QNMs of the two individual (uncoupled) systems; markers are COMSOL simulations of the full, coupled system.

## 4. Fano parameters computation

Fano theory, as established by Hugo Fano in the 60s [28], provides a framework to model interferences between a sharp resonance and a spectrally nonuniform background response [28,46,47]. One central result of the theory is the expression of the optical response of resonant systems in terms of Fano functions. For instance, the extinction cross section $\sigma_{\mathrm{ext}}$ can be written as

$$\sigma_{\mathrm{ext}}(\omega) = \sum_m \sigma_m^0 \frac{q_m^2 - 1 + 2q_m\Delta_m}{[\Delta_m^2+1](q_m^2+1)}, \quad (10)$$

where $\sigma_m^0$ is the Fano intensity characterizing the resonance strength, $q_m$ is the Fano parameter, and $\Delta_m = (\omega - \omega_m)/\gamma_m$, with $\omega_m$ and $\gamma_m$ corresponding to the real and imaginary parts of the QNM eigenfrequency $\tilde{\omega}_m = \omega_m + i\,\gamma_m/2$. In his original work [28], Fano studied only single resonances, and therefore the sum in Eq. (10) is later introduced as a rational expansion for resonant systems supporting multiple modes.

Equation (10) provides an efficient and physically transparent representation of extinction spectra involving many (potentially overlapping) resonances: the contribution of each mode to $\sigma_{\mathrm{ext}}$ is quantitatively characterized by *only* four real parameters, namely $\sigma_m^0$, $q_m$, $\omega_m$ and $\gamma_m$. The Fano factor $q_m \in [-\infty, \infty]$ provides a measure of the lineshape of the resonant contribution. When $q_m \to 0$, the profile is highly asymmetric (Fano resonance), while $|q_m| \to \infty$ corresponds to a symmetric Lorentzian lineshape. On the other hand, Fano intensity $\sigma_m^0 > 0$ quantifies the extinction strength. Finally, $\omega_m$ and $\gamma_m$ determine the frequency and linewidth of each peak.

Traditionally, the four parameters, $\sigma_m^0$ , $q_m$ , $\omega_m$ and $\gamma_m$, are extracted by fitting experimentally measured or numerically retrieved spectra [47], an approach that limits both physical insight and design intuition. Recently, closed-form expressions have been derived for $\sigma_m^0$ and $q_m$, relying on overlap integrals between the QNMs and incident fields [29]:

$$q_m = \frac{\mathrm{Re}\{\xi_m^R\} + |\xi_m^R|}{\mathrm{Im}\{\xi_m^R\}}, \quad (11)$$

and

$$\sigma_m^0 = \frac{\omega_m^2}{2I_0} \frac{|\xi_m^R|}{\gamma_m}. \quad (12)$$

Here, $\xi_m^R = \zeta_m(\omega_m, \mathbf{E}_b)\zeta_m(\omega_m, \mathbf{E}_b^*)$ , $\zeta_m(\omega, \mathbf{x}) = \iiint_{\Omega_{\mathrm{res}}} \Delta\varepsilon(\tilde{\omega}_m)\mathbf{x}(\mathbf{r}, \omega) \cdot \tilde{\mathbf{E}}_m(\mathbf{r})\, \mathrm{d}^3\mathbf{r}$ , $\Omega_{\mathrm{res}}$ is the volume of the resonator, $\mathbf{E}_b$ is the background field that excites the system, and $\Delta\varepsilon(\tilde{\omega}_m)$ is the difference of the permittivity between the resonator and the background medium, evaluated at the complex resonance frequency $\tilde{\omega}_m$. These expressions explicitly reveal how the spectral lineshape and resonance strength are governed by the spatial overlap between the QNM and incident fields. As a result, Fano theory is elevated from a heuristic tool for spectra interpretation to a predictive and quantitative framework for spectra design.

In MAN V9, we introduce two new functions in the reconstruction toolbox: one for computing the Fano parameters of Eqs. (11) and (12), and another for reconstructing the extinction spectrum using Eq. (10). These functions (**getAmplitude_and_qFano_RFA.m** and **getExtinctionFano.m**, respectively) are demonstrated across several systems in the reconstruction toolbox.

As a representative example, Fig. 7 examines a ceramic split-ring resonator with a high relative permittivity operating in the microwave regime. The split-ring supports standing-wave whispering-gallery modes that can be excited by an in-plane plane-wave. In Fig. 7(b,c), we present the computed Fano parameter $q_m$and Fano intensity $\sigma_m^0$, respectively, as the relative angle $\varphi$ between the split opening and the incident wave direction is varied, see Fig. 7(a). The QNMs are computed using the QNMEig solver.

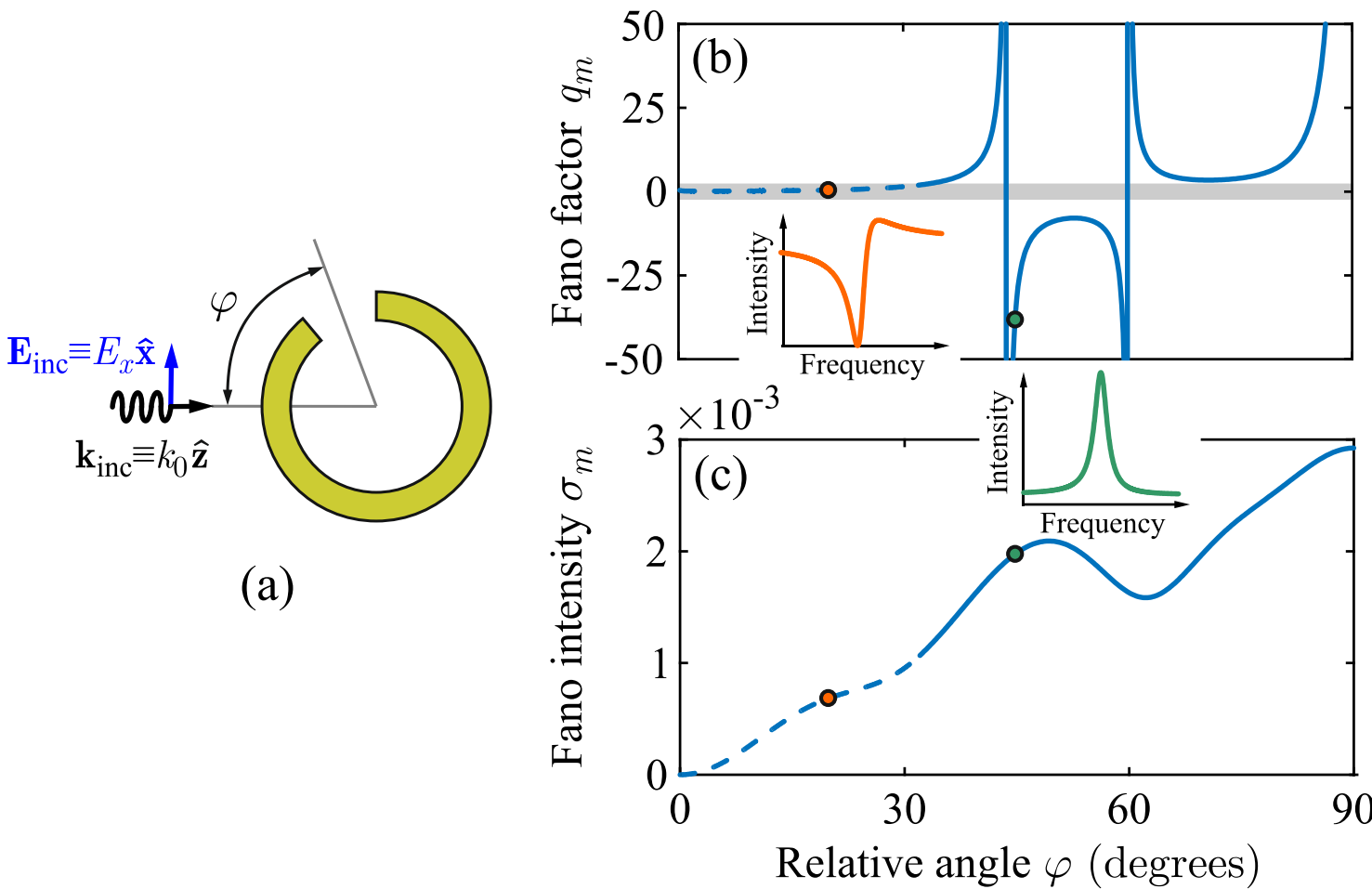

**FIG. 7**. Fano parameters computed with the reconstruction toolbox for a split-ring cavity, illuminated by an in-plane planewave. (a) Schematic of the considered split ring with 57.5 mm outer radius, 10.925 mm width, and 35° gap. (b) Fano factor and (c) Fano intensity, as calculated for the QNM retrieved by QNMEig solver and the **getAmplitude_and_qFano_RFA.m** function. Sharp asymmetric Fano lineshapes (upper inset) appear for $|q_m| < 2$ (shaded area and dashed curves) while more symmetric, Lorentzian-like lineshapes (lower inset), appear otherwise.

In general, sharp and highly asymmetric Fano resonances are observed when the Fano parameter approaches zero, whereas large values of $|q_m|$ correspond to more symmetric, Lorentzian-like line shapes [28,46]. As shown in Fig. 7(b), pronounced asymmetric resonances emerge when $q_m$ lies within the gray-shaded region, corresponding to $\varphi < 35°$, as also indicated by the dashed line and the inset on the left (orange Fano curve). Outside this region, where $|q_m| > 2$ (solid curves), the spectral response becomes increasingly symmetric (Lorentzian), as further illustrated by the inset on the right (green Lorentzian curve).

## 5. Conclusion

In this work, we have presented three major enhancements to the MAN package. First, the QNM solvers and field-reconstruction tools have been extended to natively support systems incorporating two-dimensional materials through a surface-conductivity formulation. Second, building on our recently developed rigorous coupled-QNM theory, we have introduced a new toolbox that computes coherent and dissipative coupling coefficients directly from the QNMs of the individual resonators. Unlike conventional temporal coupled-mode theory, where such parameters are often introduced phenomenologically or determined by fitting, the present approach evaluates them analytically from first principles, enabling predictive modeling of coupled photonic systems beyond the weak-coupling and high-Q regimes. Third, we have implemented a new toolbox for the direct computation of Fano parameters from overlap integrals between QNMs and incident fields. By replacing empirical fitting of spectral line shapes with analytical expressions, this toolbox provides a physically transparent and predictive interpretation of asymmetric extinction spectra.

Beyond these major additions, Version 9 includes numerous bug fixes, performance improvements, and an expanded library of COMSOL models (Fig. 1). A complete list of changes is provided in the *UpdateLOG_V9.txt* file distributed with the package. Together, these developments significantly broaden the range of nanophotonic systems accessible to MAN and further establish it as a comprehensive platform for predictive QNM-based modeling, analysis, and design.

### Declaration of competing interest

The authors declare that they have no known competing financial interests or personal relationships that could have appeared to influence the work reported in this paper.

### Data availability

Data underlying the results presented in this paper are not publicly available at this time but may be obtained from the authors upon reasonable request.

### Acknowledgements

The Authors acknowledge the contribution of O. Tsilipakos and E. E. Kriezis on the development of the 2D materials framework, and the contribution of M. Bochkarev and N. Solodovchenko on the development of the functions to compute the Fano parameters. TC and PL acknowledge financial support from the European Union's Horizon Europe Research and Innovation program under the Marie Skłodowska-Curie Actions (project QUARTE, 101148330). PL acknowledges financial support from the WHEEL (ANR-22CE24-0012-03) and UpiCO (ANR-25-CE09-4019) projects.

### References

1. P. Lalanne, W. Yan, K. Vynck, C. Sauvan, J. P. Hugonin, Light Interaction with Photonic and Plasmonic Resonances, Laser Photonics Rev **2018**(12), 1700113 (2018).
2. T. Wu, M. Gurioli, P. Lalanne, Nanoscale Light Confinement: the Q's and V's, ACS Photonics **8**(6), 1522-1538 (2021).
3. S. Both, T. Weiss, Resonant states and their role in nanophotonics, Semicond Sci Technol **37**, 013002 (2022).
4. C. Sauvan, T. Wu, R. Zarouf, E. A. Muljarov, P. Lalanne, Normalization, orthogonality and completeness of quasinormal modes of open systems: the case of electromagnetism, Opt Express **30**(5), 6846-6885 (2022).
5. T. Wu, J. L. Jaramillo, P. Lalanne, Reflections on the spatial exponential growth of electromagnetic quasinormal modes, Laser Photonics Rev **19**, 2402133 (2025).
6. T. Wu, D. Arrivault, W. Yan, P. Lalanne, Modal analysis of electromagnetic resonators: User guide for the MAN program, Comput Phys Commun **284**, 108627 (2023).
7. P. T. Leung, S. Y. Liu, K. Young, Completeness and orthogonality of quasinormal modes in leaky optical cavities, Phys Rev A **49**(4), 3057-3067 (1994).
8. E. A. Muljarov, W. Langbein, R. Zimmermann, Brillouin-Wigner perturbation theory in open electromagnetic systems, Europhys Lett **92**(5), 50010 (2010).
9. C. Sauvan, J. P. Hugonin, I. S. Maksymov, P. Lalanne, Theory of the Spontaneous Optical Emission of Nanosize Photonic and Plasmon Resonators, Phys Rev Lett **110**(23), 237401 (2013).
10. B. Vial, F. Zolla, A. Nicolet, M. Commandré, Quasimodal expansion of electromagnetic fields in open two-dimensional structures, Phys Rev A **89**(2), 023829 (2014).
11. Q. Bai, M. Perrin, C. Sauvan, J.-P. Hugonin, P. Lalanne, Efficient and intuitive method for the analysis of light scattering by a resonant nanostructure, Opt Express **21**(22), 27371-27382 (2013).
12. P. Lalanne, Mode volume of electromagnetic resonators: let us try giving credit where it is due, arXiv preprint, arXiv:2011.00218 (2020).
13. G. Demésy, A. Nicolet, B. Gralak, C. Geuzaine, C. Campos, J. E. Roman, Non-linear eigenvalue problems with GetDP and SLEPc: Eigenmode computations of frequency dispersive photonic open structures, Comput Phys Commun **257**, 107509 (2020).

14. W. Yan, R. Faggiani, P. Lalanne, Rigorous modal analysis of plasmonic nanoresonators, Phys Rev B **97**(20), 205422 (2018).
15. F. Betz, F. Binkowski, S. Burger, Rpexpand: Software for Riesz projection expansion of resonance phenomena, SoftwareX **15**, 100763 (2021).
16. F. Betz, F. Binkowski, L. Kuen, S. Burger, Version 2 – RPExpand: Software for Riesz projection expansion of resonance phenomena, SoftwareX **26**, 101694 (2024).
17. T. Wu, P. Lalanne, Rigorous electromagnetic quasinormal-mode method made easy for users, Laser Photonics Rev **20**, e03045 (2026).
18. F. J. G. de Abajo, *et al.*, Roadmap for Photonics with 2D Materials, ACS Photonics 2025, **12**(8), 3961-4095 (2025).
19. A. Liu, *et al.*, The Roadmap of 2D Materials and Devices Toward Chips. Nano-Micro Lett **16**(119), 119 (2024).
20. D. Chatzidimitriou, A. Pitilakis, E. E. Kriezis, Rigorous calculation of nonlinear parameters in graphene-comprising waveguides, J Appl Phys **118**(2), 023105 (2015).
21. Y. Gong, N. Liu, Advanced numerical methods for graphene simulation with equivalent boundary conditions: A review, Photonics **10**(7), 712 (2023).
22. T. Wu, P. Lalanne, Dissipative coupling in photonic and plasmonic resonators, Adv Photonics **7**(5), 056011 (2025).
23. H. A. Haus, *Waves and Fields in Optoelectronics*, Prentice-Hall (1984).
24. W. Suh, Z. Wang, S. Fan, Temporal coupled-mode theory and the presence of non-orthogonal modes in lossless multimode cavities, IEEE J Quantum Electron **40**(10), 1511-1518 (2004).
25. T. Christopoulos, O. Tsilipakos, E. E. Kriezis, Temporal coupled-mode theory in nonlinear resonant photonics: From basic principles to contemporary systems with 2D materials, dispersion, loss, and gain, J Appl Phys **136**(1), 011101 (2024).
26. R. Faggiani, A. Losquin, J. Yang, E. Mårsell, A. Mikkelsen, P. Lalanne, Modal analysis of the ultrafast dynamics of optical nanoresonators, ACS Photonics **4**(4), 897-904 (2017).
27. T. Wu, P. Lalanne, Exact maxwell evolution equation of resonator dynamics: temporal coupled-mode theory revisited, Opt Express **32**(12), 20904-20914 (2024).
28. U. Fano, Effects of configuration interaction on intensities and phase shifts, Phys Rev **124**(6), 1866-1878 (1961).
29. M. Bochkarev, N. Solodovchenko, K. Samusev, M. Limonov, T. Wu, P. Lalanne, Quasinormal mode as a foundational framework for all electromagnetic Fano resonances, arXiv preprint, arXiv:2412.11099 (2024).
30. T. Christopoulos, E. E. Kriezis, O. Tsilipakos, Multimode non-Hermitian framework for third harmonic generation in nonlinear photonic systems comprising two-dimensional materials, Phys Rev B **107**(3), 035413 (2023).
31. B. Majérus, E. Dremetsika, M. Lobet, L. Henrard, P. Kockaert, Electrodynamics of two-dimensional materials: Role of anisotropy, Phys Rev B **98**(12), 125419 (2018).
32. T. Christopoulos, G. Nousios, E. E. Kriezis, O. Tsilipakos, Quasinormal mode theory for multiresonant metasurfaces with superwavelength periodicity involving two-dimensional materials, Phys Rev B **110**(24), 245407 (2024).
33. P. H. Z. Cano, S. A. Amanatiadis, Z. D. Zaharis, T. V. Yioultsis, P. I. Lazaridis, N. V. Kantartzis, Accurate extraction of graphene scattering properties via a robust formulation for fundamental modes, IEEE Trans Antennas Propag **73**, 5751-5761 (2025).
34. G. W. Hanson, Dyadic Green's functions and guided surface waves for a surface conductivity model of graphene, J Appl Phys **103(6)**, 064302 (2008).

35. D. B. Lioi, P. R. Stevenson, B. T. Seymour, G. Neher, R. D. Schaller, D. J. Gosztola, R. A. Vaia, J. P. Vernon, W. J. Kennedy, Simultaneous ultrafast transmission and reflection of nanometer thick $Ti_3C_2T_x$ MXene films in the visible and near-infrared: Implications for energy storage, electromagnetic shielding, and laser systems, ACS Appl Nano Mater **3**(10), 9604-9609 (2020).
36. L. Stille, C. J. Tabert, E. J. Nicol, Optical signatures of the tunable band gap and valley-spin coupling in silicene, Phys Rev B **86**(19), 195405, (2012).
37. H.-L. Liu, C.-C. Shen, S.-H. Su, C.-L. Hsu, M.-Y. Li, and L.-J. Li, Optical properties of monolayer transition metal dichalcogenides probed by spectroscopic ellipsometry, Appl Phys Lett **105**(20), 201905 (2014).
38. T. Low, A. S. Rodin, A. Carvalho, Y. Jiang, H. Wang, F. Xia, A. H. Castro Neto, Tunable optical properties of multilayer black phosphorus thin films, Phys Rev B **90**(7), 075434 (2014).
39. P. Di Pietro, F. M. Vitucci, D. Nicoletti, L. Baldassarre, P. Calvani, R. Cava, Y. S. Hor, U. Schade, S. Lupi, Optical conductivity of bismuth-based topological insulators, Phys Rev B **86**(4), 045439 (2012).
40. J.-M. Jin, *The Finite Element Method in Electromagnetics 3rd ed.*, John Wiley & Sons (2014).
41. A. Gras, P. Lalanne, M. Duruflé, Nonuniqueness of the quasinormal mode expansion of electromagnetic lorentz dispersive materials, J Opt Soc Am A **37**(7), 1219-1228 (2020).
42. T. Wu, W. Yan, P. Lalanne, Bright plasmons with cubic nanometer mode volumes through mode hybridization, ACS Photonics **8**(1), 307-314 (2021).
43. C. Tserkezis, R. Esteban, D. O. Sigle, J. Mertens, L. O. Herrmann, J. J. Baumberg, J. Aizpurua, Hybridization of plasmonic antenna and cavity modes: Extreme optics of nanoparticle-on-mirror nanogaps, Phys Rev A **92**(5), 053811(2015).
44. I. Epstein, D. Alcaraz, Z. Huang, V. V. Pusapati, J. P. Hugonin, A. Kumar, X. M. Deputy, T. Khodkov, T. G. Rappoport, J. Y. Hong, N. M. R. Peres, J. Kong, D. R. Smith, F. H. L. Koppens, Far-field excitation of single graphene plasmon cavities with ultracompressed mode volumes, Science **368**(6496), 1219-1223 (2020).
45. M. A. Popovic, C. Manolatou, M. R. Watts, Coupling-induced resonance frequency shifts in coupled dielectric multi-cavity, Opt Express **14**(3), 1208-1222 (2006).
46. M. F. Limonov, M. V. Rybin, A. N. Poddubny, Y. S. Kivshar, Fano resonances in photonics, Nat Photonics **11**, 543-554 (2017).
47. M. F. Limonov, Fano resonance for application, Adv Opt Photonics **13**(3), 703-771 (2021).